\documentclass[aps,prb,twocolumn,superscriptaddress,longbibliography]{revtex4-2}
\usepackage{epsfig,amsopn}
\usepackage{graphicx}
\usepackage{color}
\usepackage{amsmath,amssymb}
\usepackage{enumerate}
\newcommand\bea{\begin{eqnarray}}
\newcommand\eea{\end{eqnarray}}
\newcommand\beq{\begin{equation}}
\newcommand\eeq{\end{equation}}

\def\nn{\nonumber}

\def\al{\alpha}

\def\ga{\gamma}

\def\si{\sigma}

\def\De{\Delta}
\def\dg{\dagger}

\begin{document}
\title{Four-terminal Josephson junctions: diode effects, anomalous currents and transverse currents}
\author{Bijay Kumar Sahoo}
\affiliation{School of Physics, University of Hyderabad, Prof. C. R. Rao Road, Gachibowli, Hyderabad-500046, India}
\author{ Abhiram Soori }
\email{abhirams@uohyd.ac.in}
\affiliation{School of Physics, University of Hyderabad, Prof. C. R. Rao Road, Gachibowli, Hyderabad-500046, India}

\begin{abstract}
We study a multi-terminal Josephson junction consisting of a central spin-orbit-coupled (SOC) region with an in-plane Zeeman field connected to four superconducting  terminals. This setup allows for the simultaneous measurement of both longitudinal and transverse Josephson currents in response to a phase bias and provides a platform to probe the planar Hall effect in superconducting transport. We find that the system exhibits  anomalous Josephson effect (AJE) and  Josephson diode effect (JDE) when the symmetry between opposite momentum modes is broken in SOC region. Specifically, breaking the symmetry between $k_x$ and $-k_x$ results in JDE and AJE in the longitudinal Josephson current, while breaking the symmetry between $k_y$ and $-k_y$ leads to a finite transverse Josephson current that also exhibits JDE and AJE. Transverse Josephson diode effect coefficient attains values as large as $500\%$ for realistic set of parameters.  Furthermore, for specific parameter choices, the current-phase relation  in the transverse direction supports unidirectional transport, highlighting its potential for superconducting circuit applications. Our setup offers a new route to engineering nonreciprocal superconducting transport.
\end{abstract}

\maketitle

\section{Introduction}

A nonsuperconducting material sandwiched between two superconductors (SCs) can carry an equilibrium current when a phase difference is applied between the superconductors. This phenomenon, known as the Josephson effect~\cite{JOSEPHSON1962251}, originates from bound states within the superconducting gap that mediate current flow under phase bias~\cite{FURUSAKI1999}. The current-phase relation (CPR) describes the dependence of this current on the superconducting phase difference, with its extrema termed critical currents. When both time-reversal ($TR$) and inversion ($I$) symmetries are broken in the junction, the critical currents become asymmetric, giving rise to the Josephson diode effect (JDE)—a phenomenon that has recently attracted significant theoretical and experimental attention~\cite{Ando2020,Baumgartner_2022,turini2022,Tanaka2022,soori23noneq,soori23aje,Kochan2023,Satoshi2023,Margineda2023,Nowak2024,sayan2025}.

In systems combining spin-orbit coupling (SOC) with a Zeeman field, the breaking of both $TR$ and $I$ symmetries leads to magnetochiral anisotropy, characterized by direction-dependent carrier velocities. This anisotropy underlies the emergence of JDE in spin-orbit-coupled systems~\cite{turini2022,Baumgartner_2022}. For instance, a controllable JDE was experimentally demonstrated in artificial superlattices lacking inversion symmetry~\cite{Ando2020}, and similar effects were proposed in periodically driven SNS junctions~\cite{soori23noneq} and in Floquet superconductors~\cite{soori23aje}. Experiments on InSb nanoflags showed enhanced JDE when the in-plane magnetic field is perpendicular to the current, highlighting the central role of Rashba SOC~\cite{turini2022}. Other platforms exhibiting large diode coefficients include SC–ferromagnetic insulator–SC junctions on topological insulator surfaces~\cite{Tanaka2022}, and gate-tunable 2D SNS junctions~\cite{Satoshi2023}. A diode coefficient exceeding 70\% was predicted in spin-orbit-coupled quantum dot junctions~\cite{Debika2024}, and anomalous Josephson effects were reported in semiconductor nanowires with strong SOC under magnetic fields~\cite{Yokoyama2014}. Recent theoretical efforts have proposed universal frameworks for JDE in short junctions~\cite{Margarita2022} and demonstrated its connection to the anomalous Josephson effect via gate-tunable interferometry~\cite{Reinhardt2024}. Moreover, transverse rectification and diode behavior were demonstrated in hybrid ferromagnet–superconductor devices~\cite{Strambini2022}, while symmetry requirements for anomalous Josephson currents in multilevel quantum dots were clarified in Ref.~\cite{Martin2009}. Notably, a transverse JDE was recently predicted in tilted Dirac materials~\cite{Zeng2025}.

In a parallel development, the planar Hall effect (PHE)—the generation of a transverse voltage in response to a longitudinal current—has been observed in spin-orbit-coupled two-dimensional systems subjected to an in-plane magnetic field~\cite{goldberg54,tang03,Roy10,annadi13,taskin17}. Unlike the conventional Hall effect, which originates from the Lorentz force, PHE is a consequence of the interplay between SOC and the Zeeman field~\cite{suri21,soori2021}.

Although Josephson junctions incorporating spin-orbit-coupled materials and magnetic fields have been widely studied~\cite{Wyder2002,Flensberg2016,Kochan2023}, the possibility of transverse Josephson currents—especially in multi-terminal geometries—remains largely unexplored. While transverse currents have been theoretically predicted in normal metal–superconductor junctions with interfacial SOC~\cite{Fabian2020}, and in Josephson junctions on topological insulator surfaces under in-plane fields~\cite{Oleksii2021}, there is no clear prescription for detecting such currents. Importantly, recent experiments have successfully realized multi-terminal Josephson junctions~\cite{pankra2020}, creating an opportunity to probe these transverse effects in superconducting systems.

\begin{figure}[htb]
    \centering
    \includegraphics[width=8cm]{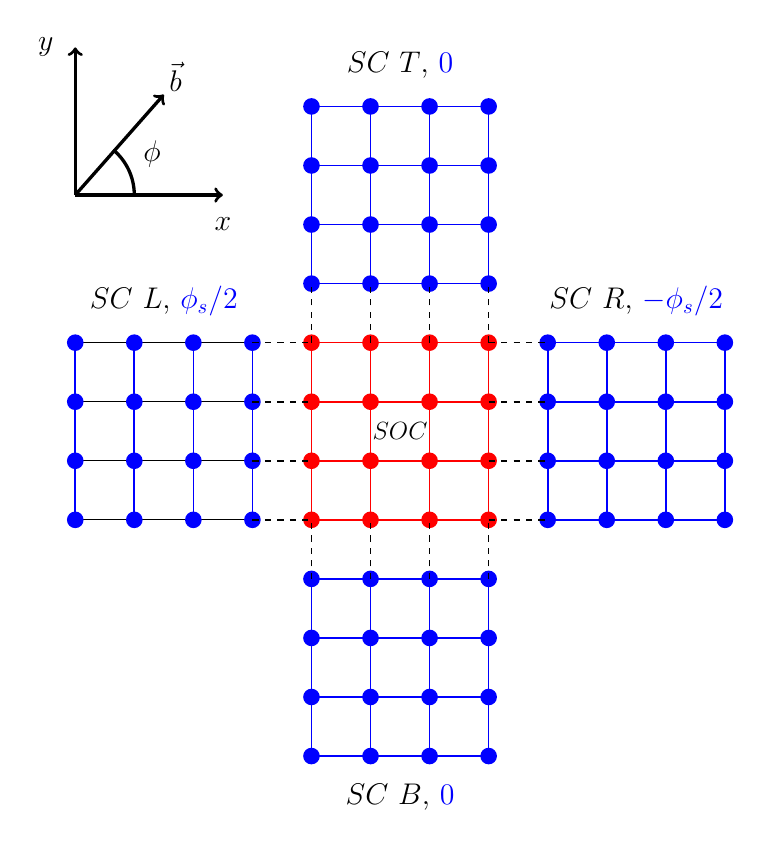}
    \caption{Schematic diagram of the proposed setup. SC on the left (right) is maintained at a phase $\phi_s/2$ ($-\phi_s/2$). SCs on top and bottom are maintained at phase $0$. An in-plane Zeeman field is applied to the central SOC region.  }
    \label{fig:schem}
\end{figure}

Motivated by these developments, we propose a Josephson junction setup involving a spin-orbit-coupled two-dimensional electron gas, where transverse currents can be measured in a multi-terminal configuration. In our design, two superconductors are connected longitudinally at opposite ends of the system, while two additional superconductors are attached in the transverse direction, as shown in Fig.~\ref{fig:schem}. While the current in the transverse terminals serves as a direct probe of transverse Josephson currents, the currents in the terminals in the longitudinal direction serve to probe the usual CPR and associated anomalous Josephson effect (AJE) and JDE. While superconducting phases on the left and the right terminals are $\phi_s/2$ and $-\phi_s/2$, the transverse terminals are maintained at zero superconducting phase. 

\section{System}
The system under study consists of a central spin-orbit-coupled (SOC) region connected to four superconducting blocks, as shown in Fig.~\ref{fig:schem}. We model the system using a tight-binding approach on a square lattice. The total Hamiltonian of the system is given by

\bea
H &=& H_L+H_{LS}+H_{SOC}+H_{SR}+H_R+H_B+H_{BS}+\nn \\ 
&& H_{ST}+ H_T, \label{eq:H} 
\eea

where $H_L$, $H_R$, $H_T$, and $H_B$ represent the Hamiltonians of the left ($L$), right ($R$), top ($T$), and bottom ($B$) superconducting blocks, respectively. The term $H_{SOC}$ describes the Hamiltonian of the central SOC region. An in-plane Zeeman field is applied to the SOC region.  The coupling between each superconducting block and the SOC region is represented by $H_{pS}$, where $p = L, R, T, B$.  The block $L$ ($R$) is maintained at phase $\phi_s/2$ ($-\phi_s/2$) and the blocks $T,B$ are maintained at phase $0$. The exact form of these Hamiltonians is detailed in Appendix~A. 

The Josephson currents $J_p$ (for $p = L, R, T, B$) at the junctions between the central SOC region and the superconducting blocks are computed as described in Appendix~B. The system is designed in such a way that when either the SOC or the Zeeman field is switched off, there is no net current that flows into the terminals- $T$ and $B$.

\section{Results}
 Using experimentally relevant parameters~\cite{Kochan2023}, we perform numerical calculations. The hopping amplitude is set to $t = 3.8$ meV, with chemical potentials in the SC, the hopping from the central region to all four SCs $t_j$ is taken to be same as $t$ and spin-orbit-coupled (SOC) regions given by $\mu_s = \mu_c = -3.6t$. The SOC strength is $\alpha = 0.5t$, the superconducting pairing potential is $\Delta = 0.06t$, and the Zeeman energy is $b = 0.1t$. The system consists of a central SOC region and four SC blocks, each modeled as a $6 \times 6$ square lattice. The currents flowing into (out of) the top/right ($J_T, J_R$) and bottom/left ($J_B, J_L$) superconductors satisfy Kirchhoff’s current law. Additionally, the relations $J_L = J_R$ ($=J_x$) and $J_T = J_B$ ($=J_y$) hold due to the symmetry of the setup and the chosen SC phase configuration.

\begin{figure}[htb]
    \centering
    \includegraphics[width=4.2cm]{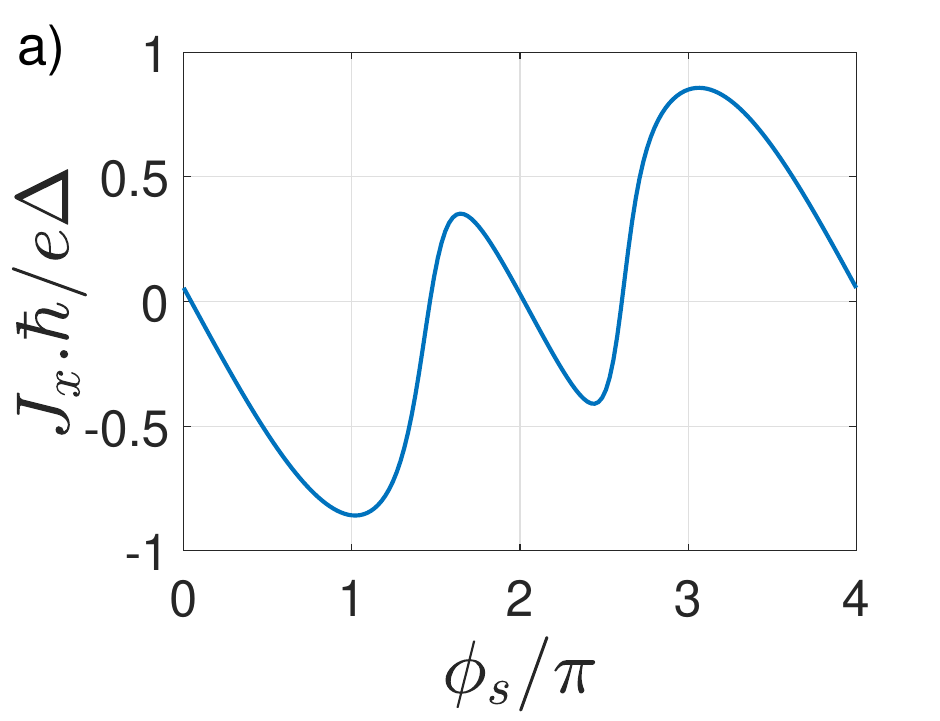}
    \includegraphics[width=4.2cm]{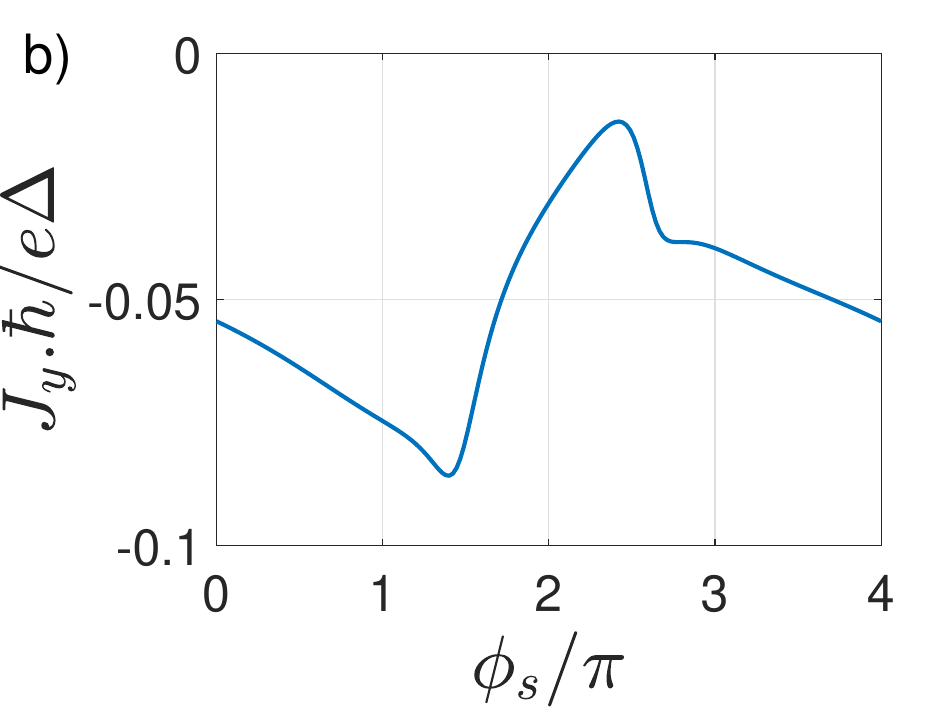}
    \includegraphics[width=4.2cm]{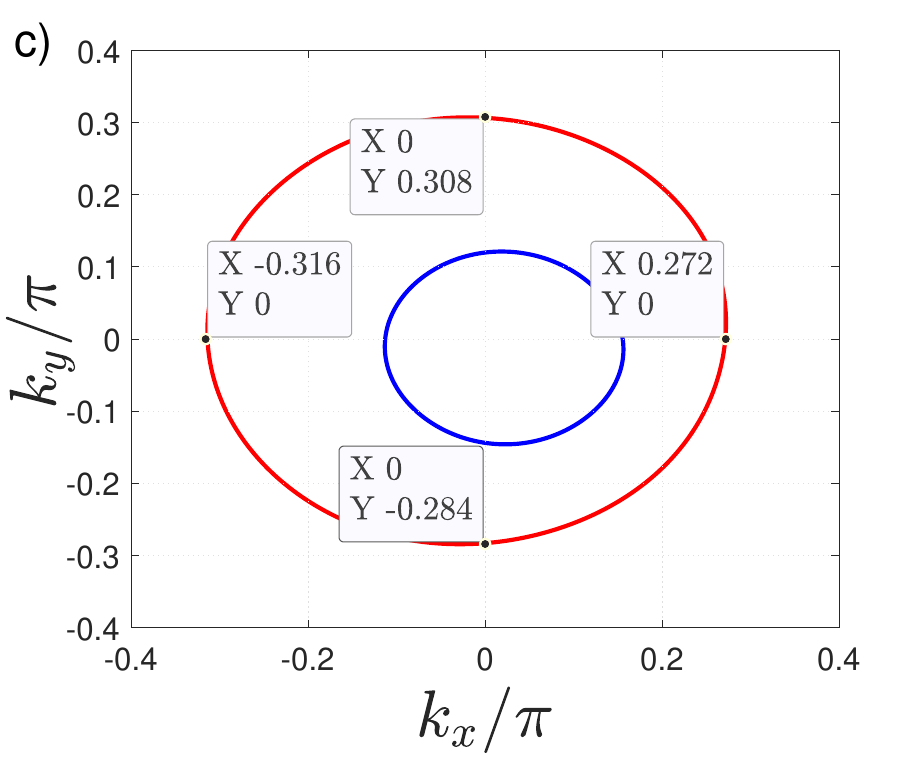}
    \includegraphics[width=4.2cm]{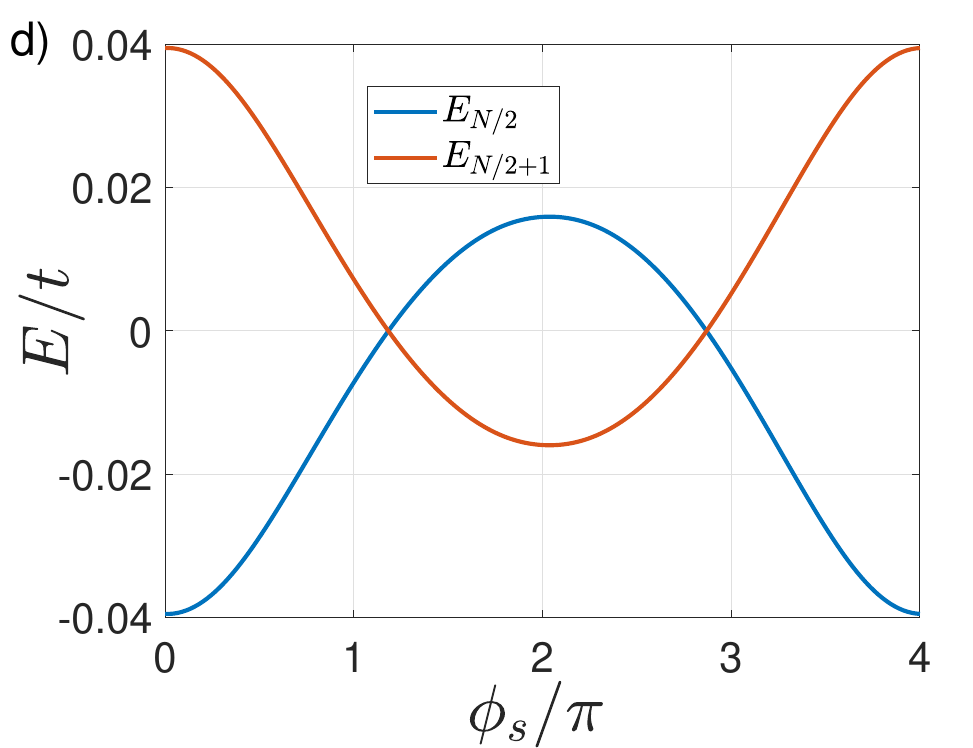}
    \caption{CPR: (a) longitudinal Josephson current, (b) transverse Josephson current, (c) Fermi surface plot, (d) Energy versus $\phi_s$. Parameters: $\al=0.5t,~b=0.1t,~\De=0.06t,~\phi=\pi/3,~\mu_c=\mu_s= -3.6t$. The number of sites in $x$ and $y$-directions for SC ($L_s^x,~L_s^y$) and SOC ($L_c^x,~L_c^y$) regions are  $L_s^x=L_s^y=L_c^x=L_c^y=6$.}
    \label{fig:cpr}
\end{figure}

In Fig.~\ref{fig:cpr}, the CPRs for longitudinal and transverse currents ($J_x$ and $J_y$) are plotted for a Zeeman field of strength $b = 0.1t$ oriented at an angle $\phi = \pi/3$ with respect to the $x$-axis. Due to the involvement of three SC phases ($\phi_s/2, 0, -\phi_s/2$), the CPRs exhibit a $4\pi$-periodicity. To better understand  the $4\pi$-periodicity of the Josephson current, we plot the last occupied energy level ($E_{N/2}$) and the first unoccupied level ($E_{N/2+1}$) versus $\phi_s$ in Fig.~\ref{fig:cpr} (d). Here we can see that both these energy levels show $4\pi$-periodicity with $\phi_s$, for which CPR also exhibits $4\pi$-periodicity. Both the longitudinal and the transverse currents display AJE and JDE.
The Hamiltonian for the central SOC region in momentum space is given by  
\begin{equation}
H(\mathbf{k}) = \epsilon_k \sigma_0 + \alpha (\sin k_y \sigma_x - \sin k_x \sigma_y) + b (\cos \phi \sigma_x + \sin \phi \sigma_y),
\end{equation}
where $\epsilon_k = -2t (\cos k_x + \cos k_y) - \mu_c$. For $\phi \neq \pi/2, 3\pi/2$, the asymmetry between $k_y$ and $-k_y$ leads to a finite transverse current . Similarly, for $\phi \neq 0, \pi$, the asymmetry between $k_x$ and $-k_x$ results in JDE and AJE~\cite{soori24bam,soori25mixed}. In other words, for every $k_x$ at any given energy, $-k_x$ is not present at the same energy in the dispersion of the central SOC region, which leads to JDE and AJE. In similar fashion, absence of $-k_y$ for a given $k_y$ at fixed energy in the dispersion of the central SOC region leads to transverse Josephson currents and diode effect in it. 
We can see the asymmetries associated with $k_x$ and $-k_x$ ($k_y$ and $-k_y$) in the Fermi surface plot of SOC region in Fig.~\ref{fig:cpr} (c) for $\phi=\pi/3$. Notably, Fig.~\ref{fig:cpr}(b) reveals that the transverse Josephson current is unidirectional for this parameter set.

\begin{figure}
    \centering
    \includegraphics[width=4.2cm]{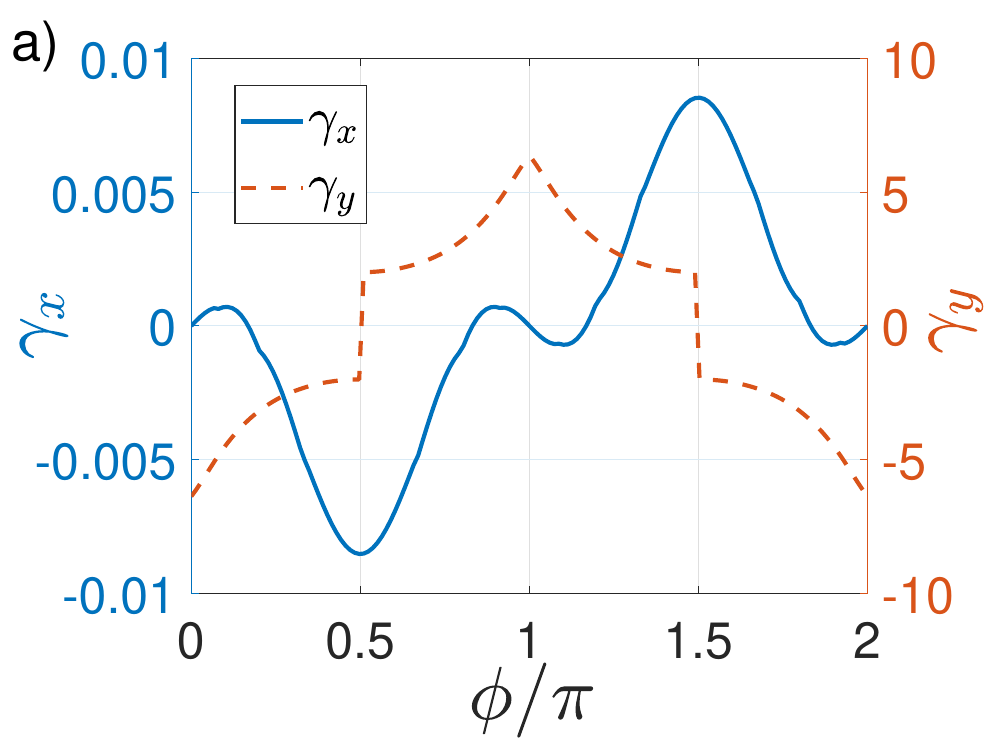}
    \includegraphics[width=4.2cm]{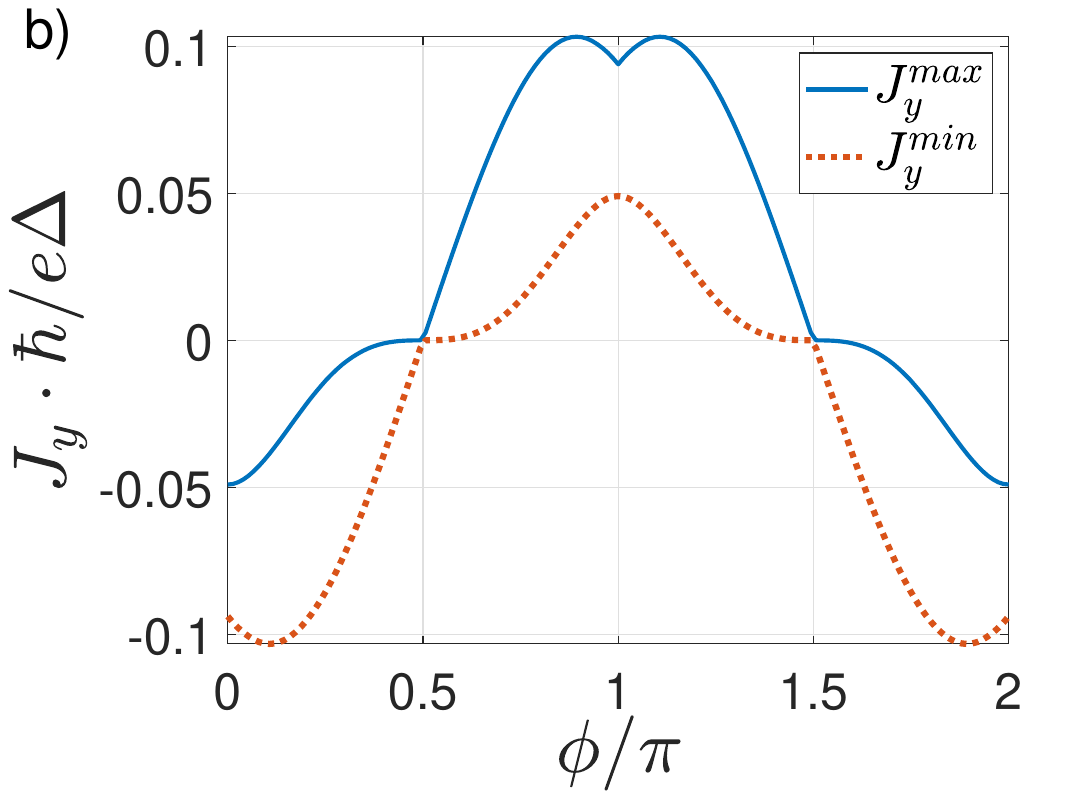}
    \caption{a) Diode effects coefficient for longitudinal (transverse) CPR $\ga_x$ ($\ga_y$) versus $\phi$ -the angle made by the in-plane Zeeman field with $x$-direction. (b) Transverse critical currents versus magnetization angle $\phi$. Other parameters are the same as in Fig.~\ref{fig:cpr}.}
    \label{fig:coef}
\end{figure}

To quantify the diode effect, we define the longitudinal and transverse diode coefficients as  
\begin{equation}
\gamma_d = \frac{2(J_d^{\max} + J_d^{\min})}{J_d^{\max} - J_d^{\min}}, \quad d = x, y.
\end{equation}
In Fig.~\ref{fig:coef}~(a) we plot $\gamma_x$ and $\gamma_y$ as functions of $\phi$, the angle in-plane Zeeman field makes with $\hat x$. As expected, the longitudinal diode effect vanishes at $\phi = 0, \pi$, while the transverse diode effect disappears at $\phi = \pi/2, 3\pi/2$. Additionally, Fig.~\ref{fig:coef}(b) shows the variation of transverse critical currents with $\phi$, demonstrating that the transverse current remains unidirectional and vanishes at $\phi = \pi/2, 3\pi/2$. In the range $\phi \in [\pi/2, \pi]$, the increasing asymmetry between $k_y$ and $-k_y$ leads to a monotonic dependence of the transverse diode coefficient on $\phi$. A similar argument explains the overall behaviour of the transverse diode effect coefficient across the entire range of $\phi$.

\begin{figure}
    \centering
    \includegraphics[width=6cm]{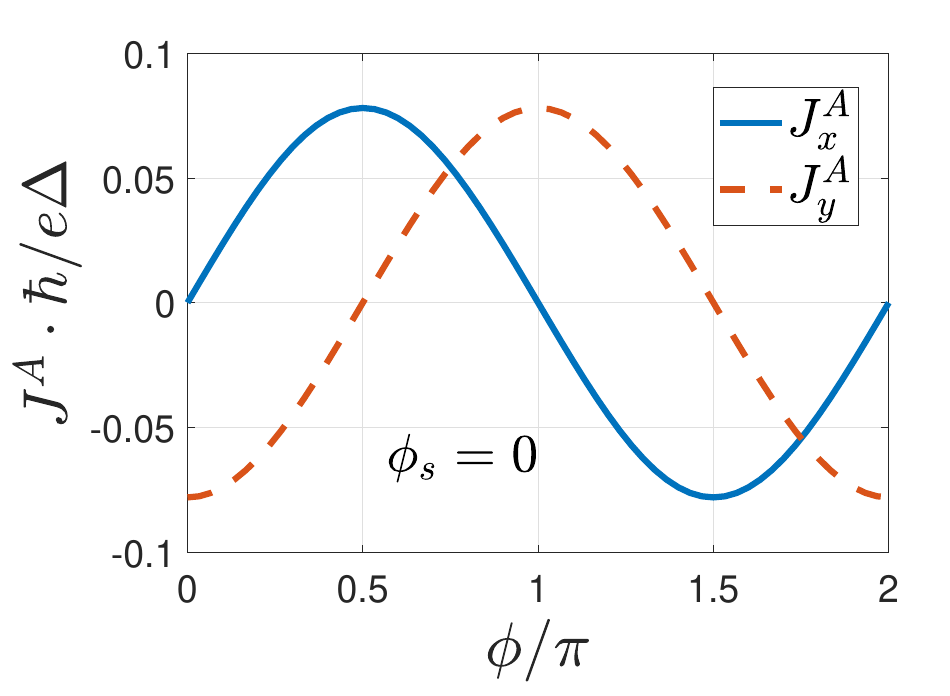}
    \caption{Longitudinal and Transverse anomalous Josephson currents ($J_x^A$ and $J_y^A$) versus magnetization angle $\phi$. Other parameters are the same as in Fig.~\ref{fig:cpr}. }
    \label{fig:aje}
\end{figure}

We plot the longitudinal anomalous Josephson current ($J_x^A$) and transverse anomalous Josephson current ($J_y^A$) versus the magnetization angle $\phi$ in Fig.~\ref{fig:aje} for the same set of parameters as in Fig.~\ref{fig:coef}. Here, we can see that $J_x^A$ and $J_y^A$ vary sinusoidally versus the magnetization angle $\phi$, which is a direct consequence of the combination of SOC and the Zeeman field. And $J_x^A$ is zero for $\phi=0,\pi$, because at these values of $\phi$, the $y$-component of the Zeeman field is zero and hence the symmetry between $k_x$ and $-k_x$ modes is not broken. Similarly, the transverse current is zero for $\phi=\pi/2,3\pi/2$ as the $x$ component of the Zeeman field is zero at these values of $\phi$ and the symmetry between $k_y$ and $-k_y$ is not broken. Notably, if the curve for $J_x^A$ is shifted by $\pi/2$ along $\phi$-axis, one gets $J_y^A$ curve. This is because, the system is symmetric under $\pi/2$ rotation in absence of $\vec b$-field and $\vec b$-field alone dictates the direction and magnitude of the anomalous Josephson current since the phase bias is absent. 

\begin{figure}
    \centering
    \includegraphics[width=4.2cm]{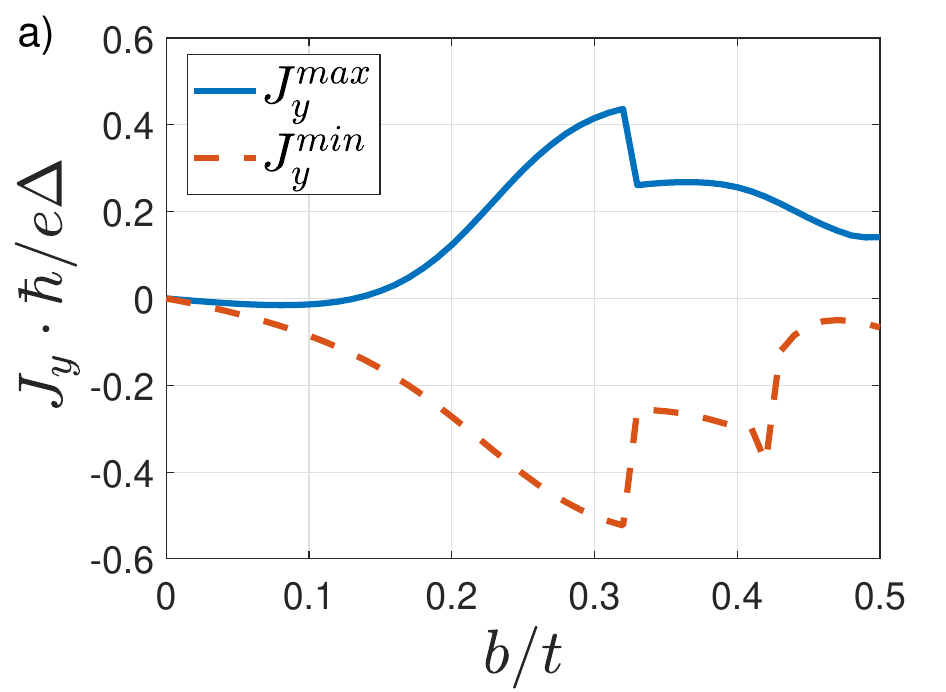}
    \includegraphics[width=4.2cm]{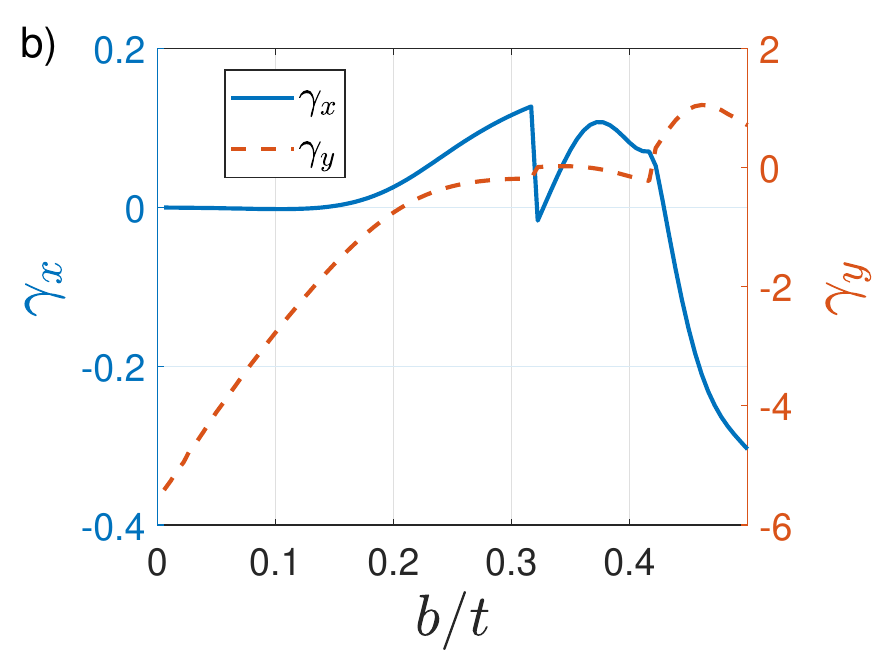}
    \caption{(a)Transverse critical Josephson currents $J_y^{max}$ and ${J_y^{min}}$ versus the Zeeman energy $b$, (b) Longitudinal (transverse) diode effect coefficient $\gamma_x$ ($\gamma_y$) versus $b$. Both the figures are for $\phi=\pi/4$ and other parameters are the same as in Fig.~\ref{fig:cpr}.}
    \label{fig:b}
\end{figure}

Next, we examine the behavior of transverse critical Josephson currents with the Zeeman energy $b$ in Fig.~\ref{fig:b}~(a) for $\phi=\pi/4$, keeping the other parameters unchanged. We can see from this figure that in the range $0<b< 0.13t$, both $J_y^{max}$ and $J_y^{min}$ are of the same sign, which means that the transverse Josephson current is unidirectional in this range. This behavior changes as the system size is increased. In Fig.~\ref{fig:b}~(b), we plot the diode effect coefficients $\gamma_x$ and $\gamma_y$ versus $b$. As $b$ increases, the longitudinal diode effect coefficient increases. On the other hand, the transverse diode effect coefficient is large for small $b$. However, the transverse critical currents are small in magnitude for small $b$, making the diode effect less prominent in this limit. 

\begin{figure}
    \centering
    \includegraphics[width=4.2cm]{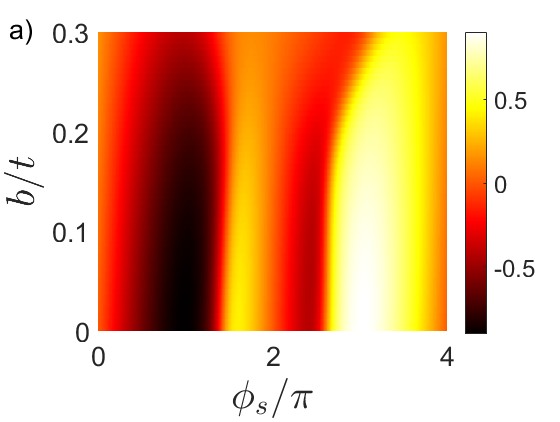}
    \includegraphics[width=4.2cm]{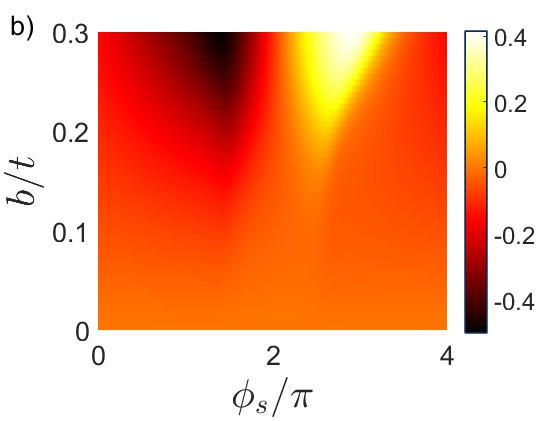}
    \caption{a) Longitudinal Josephson current $J_x$ versus superconducting phase difference $(\phi_s)$ and the Zeeman energy $b$, b) Transverse Josephson current $J_y$ versus $\phi_s$ and $b$. All figures are for $\phi=\pi/4$ and other parameters are the same as in Fig.~\ref{fig:cpr}.}
    \label{fig:b-phis}
\end{figure}

In Fig. \ref{fig:b-phis}~(a,b), we plot the longitudinal and transverse Josephson currents versus superconducting phase difference $\phi_s$ and the Zeeman energy $b$. In Fig. \ref{fig:b-phis}~(b), for a fixed value of $\phi_s$, the value of the transverse Josephson current increases with $b$. Because,  the asymmetry associated with $+k_y$ and $-k_y$ increases with $b$, resulting in an increased value of the transverse Josephson current.

\begin{figure}
    \centering
    \includegraphics[width=4.2cm]{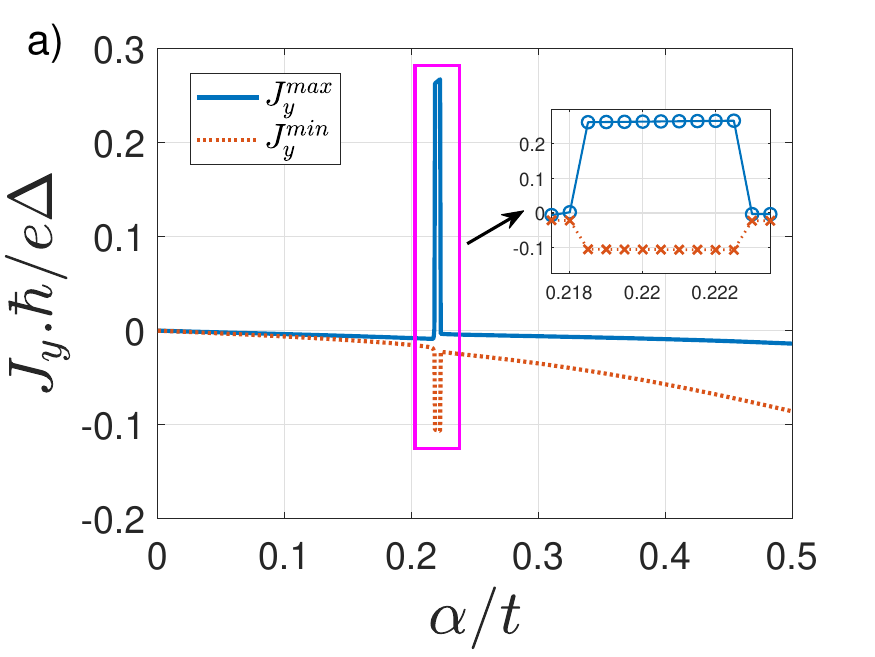}
    \includegraphics[width=4.2cm]{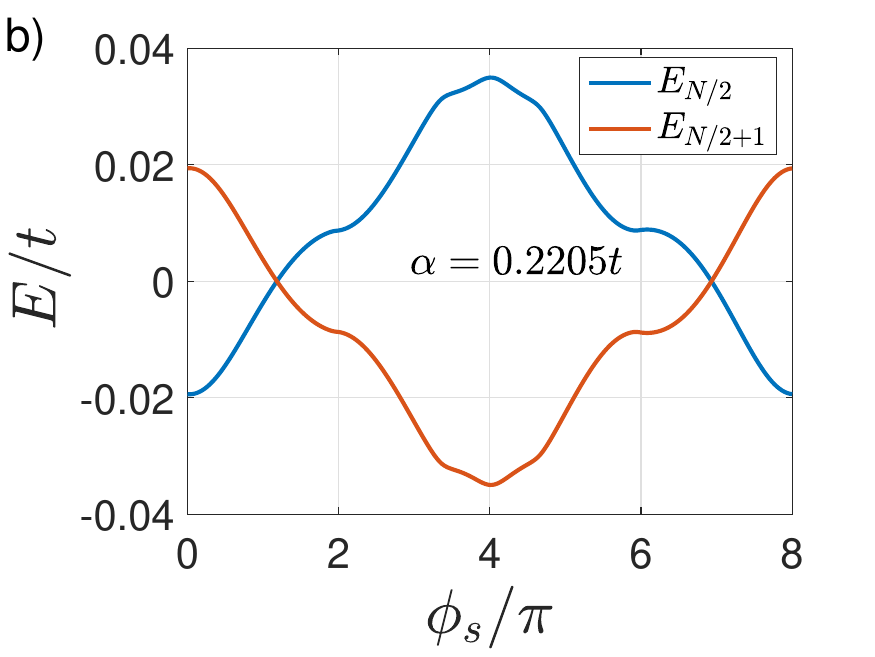}
    \includegraphics[width=4.2cm]{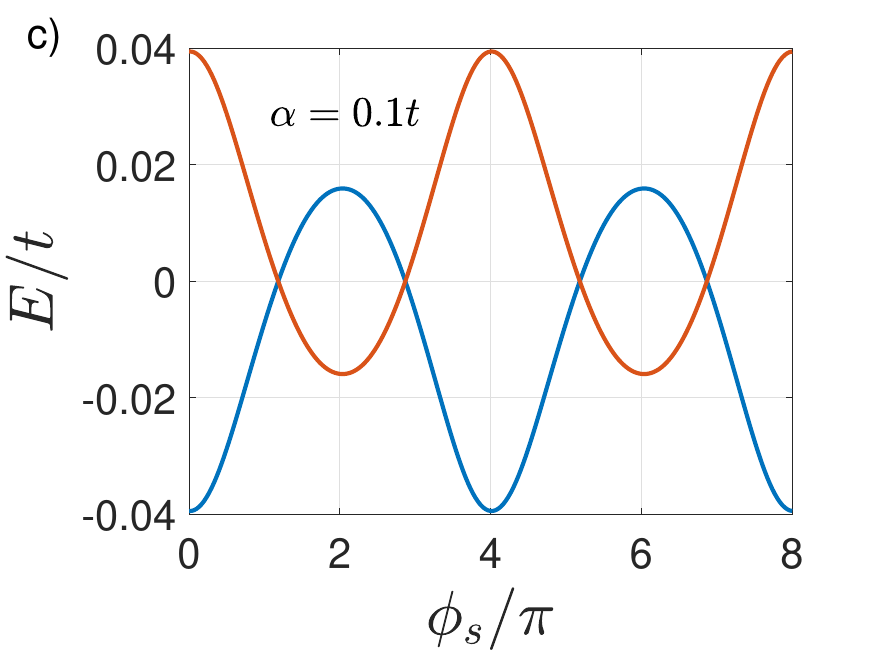}
    \includegraphics[width=4.3cm]{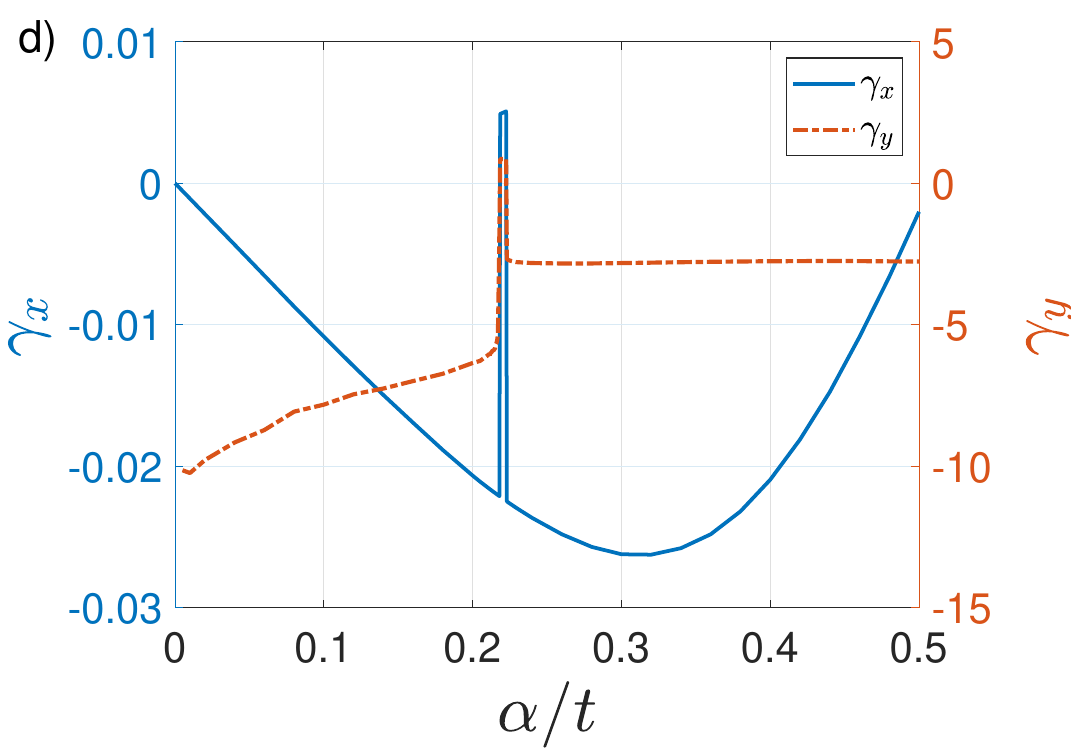}
    \caption{(a) Transverse critical Josephson currents $J_y^{max}$ and ${J_y^{min}}$ versus the SOC strength $\alpha$. Energy versus superconducting phase difference $\phi_s$ in Fig. (b) and Fig. (c) for $\alpha=0.2205t$ and $\alpha=0.1t$, respectively, (d) Longitudinal (transverse) diode effect coefficient $\gamma_x$ ($\gamma_y$) versus $\alpha$. All figures are for $\phi=\pi/4$ and other parameters are the same as in Fig.~\ref{fig:cpr}.} 
    \label{fig:Jy-alpha}
\end{figure}
Next, we check the behavior of the transverse critical Josephson currents with the spin-orbit coupling strength $\alpha$ in Fig.~\ref{fig:Jy-alpha}~(a) for $\phi=\pi/4$ and other parameters are the same as in Fig.~\ref{fig:cpr}. We can see from the figure that both $J_y^{max}$ and $J_y^{min}$ are of the same sign except for a small range of $\alpha$ values ($\alpha \in [0.218t,0.223t]$) in the middle. We show the zoomed version of that region in the inset of this figure. There is a sudden jump in the current value, because the CPR exhibits $8\pi$-periodicity for that $\alpha$ range. For more detail, we plot the last occupied energy level $E_{N/2}$ and first unoccupied energy level $E_{N/2+1}$ versus $\phi_s$ in Fig.~\ref{fig:Jy-alpha}~(b) and Fig.~\ref{fig:Jy-alpha}~(c) for $\alpha=0.2205t$ and $\alpha=0.1t$ respectively.   We can see that in Fig.~\ref{fig:Jy-alpha}~(b), the energy levels are $8\pi$ periodic, while in Fig.~\ref{fig:Jy-alpha}~(c) $4\pi$ periodic. For this reason, we see a sudden jump in the Josephson current for the above-mentioned $\alpha$ values. In Fig.~\ref{fig:Jy-alpha}~(d) we plot the coefficients of the longitudinal and transverse diode effect versus $\alpha$ for $\phi=\pi/4$ and keep other parameters the same as in Fig.~\ref{fig:cpr}. We can see that in Fig.~\ref{fig:Jy-alpha}~(d), both $\gamma_x$ and $\gamma_y$ show a sudden jump for that small range of $\alpha$, and an explanation similar to the one for transverse critical currents in Fig.~\ref{fig:Jy-alpha}~(a) holds here too.

In contrast with the results of Ref.~\cite{Kochan2023} in which diode effect coefficients of up to $30\%$ are reported, we obtain a maximum (longitudinal) diode effect coefficient of about $3\%$. Qualitatively, the difference between the two systems is that in Ref.~\cite{Kochan2023}, the setup is a two-terminal Josephson junction with translational invariance along $\hat y$-direction, whereas the setup studied by us is a 4-terminal setup with two terminals in the transverse direction. Transverse diode effect has been studied in tilted Dirac systems, where a diode effect coefficient of $100\%$ has been obtained~\cite{Zeng2025}. In contrast, we find that a transverse diode effect coefficient of $500\%$ is possible in our setup. 

\section{Summary and Conclusion}
A controlled SC phase difference can be experimentally applied in two-terminal Josephson junctions~\cite{frolov2004,glick2018}. With experimental advances in multi-terminal Josephson junctions~\cite{pankra2020,Gupta2023}, the setup we propose should be within experimental reach.

To summarize, we have proposed a setup consisting of a SOC region connected to four SC terminals in a configuration that allows probing the planar Hall effect in Josephson currents by analyzing the response to a phase bias in the longitudinal direction. This setup enables simultaneous measurement of both longitudinal and transverse Josephson currents. We find that when the symmetry between $k_x$ and $-k_x$ modes is broken, the system exhibits both the JDE and the AJE. Similarly, when the symmetry between $k_y$ and $-k_y$ modes is broken, a finite transverse Josephson current emerges, with both AJE and JDE appearing in the transverse response. Notably, for certain parameter choices, the CPR exhibits unidirectional transport in the transverse direction, highlighting the potential for practical applications in superconducting devices.

\section*{Acknowledgements}
AS and BKS  thank Science and Engineering Research Board (now Anusandhan National Research Foundation) Core Research grant (CRG/2022/004311) for financial support.  AS thanks University of Hyderabad  for funding through Institute of Eminence Professional Development Fund. BKS thanks the Ministry of Social Justice and Empowerment, Government of India, for a fellowship through NFOBC.
\bibliography{ref_scphe}

\begin{widetext}

\appendix
\section{Hamiltonian }
Different terms that make up the Hamiltonian in eq.~\eqref{eq:H} are given below. 
\bea   
H_L &=& \sum_{n_x=1}^{L_s^x-1} \sum_{n_y=L_s^y+1}^{L_{sc}^y} \Big[ -t (\Psi_{n_x+1,n_y}^{\dg}\tau_z \Psi_{n_x,n_y}+h.c.)\Big] +\sum_{n_x=1}^{L_s^x} \sum_{n_y=L_s^y+1}^{L_{sc}^y-1}\Big[ -t (\Psi_{n_x,n_y+1}^{\dg}\tau_z \Psi_{n_x,n_y}+h.c.) \Big]\nn \\
&& -\mu_s \sum_{n_x=1}^{L_s^x} \sum_{n_y=L_s^y+1}^{L_{sc}^y}\Psi_{n_x,n_y}^{\dg}\tau_z \Psi_{n_x,n_y} -\De \sum_{n_x=1}^{L_s^x} \sum_{n_y=L_s^y+1}^{L_{sc}^y}\Psi_{n_x,n_y}^{\dg} (cos\phi_l \tau_y\si_y+sin\phi_l \tau_x\si_y) \Psi_{n_x,n_y}, \nn \\
H_{LS} &=& -t_j \sum_{n_y=L_s^y+1}^{L_{sc}^y}(\Psi_{L_s+1,n_y}^{\dg} \tau_z \Psi_{L_s,n_y}+h.c.), \nn \\
H_{SOC} &=& \sum_{n_x=L_s^x+1}^{L_{sc}^x-1} \sum_{n_y=L_s^y+1}^{L_{sc}^y} \Big[ -t (\Psi_{n_x+1,n_y}^{\dg}\tau_z \Psi_{n_x,n_y}+h.c.)\Big] +\sum_{n_x=L_s^x+1}^{L_{sc}^x} \sum_{n_y=L_s^y+1}^{L_{sc}^y-1}\Big[ -t (\Psi_{n_x,n_y+1}^{\dg}\tau_z \Psi_{n_x,n_y}+h.c.) \Big]\nn \\
&& -\mu_c \sum_{n_x=L_s^x+1}^{L_{sc}^x}\sum_{n_y=L_s^y+1}^{L_{sc}^y}\Psi_{n_x,n_y}^{\dg}\tau_z \Psi_{n_x,n_y} + \frac{\al}{2} \sum_{n_x=L_s^x+1}^{L_{sc}^x}\sum_{n_y=L_s^y+1}^{L_{sc}^y-1}(i\Psi_{n_x,n_y+1}^{\dg}\tau_0\si_x \Psi_{n_x,n_y}+h.c.)\nn \\
&&-\frac{\al}{2}\sum_{n_x=L_s^x+1}^{L_{sc}^x-1} \sum_{n_y=L_s^y+1}^{L_{sc}^y}(i\Psi_{n_x+1,n_y}^{\dg}\tau_z\si_y \Psi_{n_x,n_y}+h.c.) +b \sum_{n_x=L_s^x+1}^{L_{sc}^x} \sum_{n_y=L_s^y+1}^{L_{sc}^y}\Big[\Psi_{n_x,n_y}^{\dg}(cos\phi \tau_z\si_x+sin\phi \tau_0\si_y)\Psi_{n_x,n_y}\Big],\nn \\
H_{SR} &=& -t_j \sum_{n_y=L_s^y+1}^{L_{sc}^y}(\Psi_{L_{sc}+1,n_y}^{\dg} \tau_z \Psi_{L_{sc},n_y}+h.c.), \nn \\
H_R &=& \sum_{n_x=L_{sc}^x+1}^{L_{scs}^x-1} \sum_{n_y=L_s^y+1}^{L_{sc}^y} \Big[ -t (\Psi_{n_x+1,n_y}^{\dg}\tau_z \Psi_{n_x,n_y}+h.c.)\Big] +\sum_{n_x=L_{sc}^x+1}^{L_{scs}^x} \sum_{n_y=L_s^y+1}^{L_{sc}^y-1}\Big[ -t (\Psi_{n_x,n_y+1}^{\dg}\tau_z \Psi_{n_x,n_y}+h.c.) \Big] \nn \\
&& -\mu_s \sum_{n_x=L_{sc}^x+1}^{L_{scs}^x}\sum_{n_y=L_s^y+1}^{L_{sc}^y}\Psi_{n_x,n_y}^{\dg}\tau_z \Psi_{n_x,n_y} -\De \sum_{n_x=L_{sc}^x+1}^{L_{scs}^x}\sum_{n_y=L_s^y+1}^{L_{sc}^y}\Psi_{n_x,n_y}^{\dg} (cos\phi_r \tau_y\si_y+sin\phi_r \tau_x\si_y) \Psi_{n_x,n_y}, \nn \\
H_{BS} &=& -t_j\sum_{n_x=L_{s}^x+1}^{L_{sc}^x} (\Psi_{n_x,L_{s}+1}^{\dg} \tau_z \Psi_{n_x,L_{s}}+h.c.), \nn \\
H_B &=& \sum_{n_x=L_{s}^x+1}^{L_{sc}^x-1} \sum_{n_y=1}^{L_{s}^y}\Big[ -t (\Psi_{n_x+1,n_y}^{\dg}\tau_z \Psi_{n_x,n_y}+h.c.)\Big]+\sum_{n_x=L_{s}^x}^{L_{sc}^x} \sum_{n_y=1}^{L_{s}^y-1}\Big[ -t (\Psi_{n_x,n_y+1}^{\dg}\tau_z \Psi_{n_x,n_y}+h.c.) \Big]\nn \\ && -\mu_s \sum_{n_x=L_{s}^x+1}^{L_{sc}^x}\sum_{n_y=1}^{L_{s}^y}\Psi_{n_x,n_y}^{\dg}\tau_z \Psi_{n_x,n_y}-\De \sum_{n_x=L_{s}^x+1}^{L_{sc}^x}\sum_{n_y=1}^{L_{s}^y}\Psi_{n_x,n_y}^{\dg} \tau_y \si_y \Psi_{n_x,n_y},\nn \\
H_{ST} &=&-t_j \sum_{n_x=L_{s}^x+1}^{L_{sc}^x} (\Psi_{n_x,L_{sc}+1}^{\dg} \tau_z \Psi_{n_x,L_{sc}}+h.c.), \nn \\
H_T &=& \sum_{n_x=L_{s}^x+1}^{L_{sc}^x-1} \sum_{n_y=L_{sc}^y+1}^{L_{scs}^y} \Big[ -t (\Psi_{n_x+1,n_y}^{\dg}\tau_z \Psi_{n_x,n_y}+h.c.)\Big]+\sum_{n_x=L_{s}^x}^{L_{sc}^x} \sum_{n_y=L_{sc}^y+1}^{L_{scs}^y-1}\Big[ -t (\Psi_{n_x,n_y+1}^{\dg}\tau_z \Psi_{n_x,n_y}+h.c.) \Big]\nn \\ && -\mu_s \sum_{n_x=L_{s}^x+1}^{L_{sc}^x}\sum_{n_y=L_{sc}^y+1}^{L_{scs}^y}\Psi_{n_x,n_y}^{\dg}\tau_z \Psi_{n_x,n_y}-\De \sum_{n_x=L_{s}^x+1}^{L_{sc}^x}\sum_{n_y=L_{sc}^y+1}^{L_{scs}^y}\Psi_{n_x,n_y}^{\dg} \tau_y \si_y \Psi_{n_x,n_y}, \nn \\
\eea

Here, $L_{sc}^d = L_s^d + L_c^d$ and $L_{scs}^d = 2L_s^d + L_c^d$, where $d = x, y$. The quantities $L_s^d$ and $L_c^d$ represent the number of sites in the superconducting (SC) and spin-orbit-coupled (SOC) regions along direction $d$. The spinor $\Psi_{n_x,n_y}$ is defined as  
\begin{equation}
\Psi_{n_x,n_y} = \begin{bmatrix} c_{n_x,n_y,\uparrow} & c_{n_x,n_y,\downarrow} & c_{n_x,n_y,\uparrow}^\dagger & c_{n_x,n_y,\downarrow}^\dagger \end{bmatrix}^T,
\end{equation}  
where $c_{n_x,n_y,\sigma}$ is the annihilation operator for an electron with spin $\sigma$ at site $(n_x,n_y)$. The Pauli matrices $\tau_{x,y,z}$ and $\sigma_{x,y,z}$ act on the particle-hole and spin spaces, respectively. Other parameters include:  

\begin{itemize}
  \item $t$: hopping strength in the SC and SOC regions,  
  \item $t_j$: hopping strength connecting the SC and SOC regions, chosen to be equal to $t$ in our work,  
  \item $\Delta$: superconducting pairing strength,  
  \item $\alpha$: spin-orbit coupling strength,  
  \item $b$: Zeeman energy,  
  \item $\phi$: angle between the Zeeman field and the $x$-axis,  
  \item $\mu_s$ ($\mu_c$): chemical potential in the SC (SOC) region.  
\end{itemize}

\section{Josephson currents}
The Josephson current is given by the sum of currents carried by all occupied states~\cite{Soori_2020}. Since charge is conserved in the SOC region, the charge current operators at  four junctions between the SOC region and superconductors are:  

\begin{align}  
\hat{J_L} &= \frac{iet_j}{\hbar} \sum_{n_y=L_{s}^y+1}^{L_{sc}^y} (\Psi_{L_s+1,n_y}^\dagger \Psi_{L_s,n_y} - \text{h.c.}) \\
\hat{J_R} &= \frac{iet_j}{\hbar} \sum_{n_y=L_{s}^y+1}^{L_{sc}^y} (\Psi_{L_{sc}+1,n_y}^\dagger \Psi_{L_{sc},n_y} - \text{h.c.}) \\
\hat{J_B} &= \frac{iet_j}{\hbar} \sum_{n_x=L_{s}^x+1}^{L_{sc}^x} (\Psi_{n_x,L_s+1}^\dagger \Psi_{n_x,L_s} - \text{h.c.}) \\
\hat{J_T} &= \frac{iet_j}{\hbar} \sum_{n_x=L_{s}^x+1}^{L_{sc}^x} (\Psi_{n_x,L_{sc}+1}^\dagger \Psi_{n_x,L_{sc}} - \text{h.c.})  
\end{align}  

For each $\phi_s$, the eigenstates and eigenenergies $(|u_j\rangle, E_j)$ of the Hamiltonian are obtained via numerical diagonalization. We assume that all negative energy states are occupied while positive energy states remain empty at $\phi_s \to 0^+$. At other values of $\phi_s$, the occupied states are tracked by incrementally evolving them from the initially filled states.  

The total Josephson current is given by  
\begin{equation}  
J_p = \sum_j \langle u_j | \hat{J_p} | u_j \rangle,  
\end{equation}  
where $p = L, R, T, B$ and $j$ is summed over occupied states.

In the limit of superconducting leads becoming semi-infinite in length, the contribution to Josephson current from the states outside the superconducting gap tends to zero and all the Josephson current is carried by the subgap Andreev bound states~\cite{FURUSAKI1999}. When a voltage bias is applied, the states outside the gap carry dissipative current in a system with semi-infinite leads. The system considered here is a finite, closed system and the expectation values of the current are taken using eigenstates. Hence, our calculations corresponds to equilibrium non-dissipative Josephson currents. 

\end{widetext}

\end{document}